\documentstyle [12pt] {article}
\input epsf

\begin{document}
\oddsidemargin -.375in
\begin{flushright}
%ISU-NP-04-15\\
%KKU-NP-05-02\\
%March 2005\
\end{flushright}
\vspace {.5in}
\begin{center}
{\Large\bf Superradiance and subradiance in the electromagnetic
radiative decay of $X(3872)$\\} \vspace{.5in} {\bf A. Abd El-Hady
${}^{} \footnote{Permanent address : Physics Department, Faculty
of Science, Zagazig University, Zagazig, Egypt}$\\} \vspace{.1in}
{\it Physics Department, Faculty of Science, King Khalid
University, Abha 9004, Saudi Arabia}\\
\vskip .5in
%{\bf Abstract}

\bf{\today}

\end{center}

\begin{abstract}

It is pointed out that if the molecular interpretation of the
recently observed resonance $X(3872)$ is valid, then nature may
have prepared a good laboratory for us to examine the phenomenon
of superradiance and subradiance of Dicke. The superradiance and
supradiance factors are evaluated and the effects on the
electromagnetic radiative decay of $X(3872)$ are discussed. Our
results using coordinate space representation is similar to the
momentum space results of Voloshin.
\end{abstract}

\newpage

%\twocolumn

%

\section{Introduction\label{intro}}
The narrow resonance $X(3872)$ observed by the Belle Collaboration
\cite{x(3872)_Belle} in the decay channel $X \rightarrow \pi^+
\pi^- J/\psi$ has attracted a great experimental and theoretical
interest. It has been experimentally confirmed by CDF
\cite{x(3872)_CDF}, D\O \cite{x(3872)_D0}, and BabBar
\cite{x(3872)_BaBar} Collaborations. Theoretically, different
interpretations have been suggested to clarify the nature of the
$X(3872)$ resonance. Charmonium options were considered for
$X(3872)$ \cite{charmonium_options} while coupled channel effects
were discussed in \cite{coupled_channels}. Exotic interpretations
like a diqurak-antidiqurak (tetraqurak) state
$[cq][\bar{c}\bar{q}]$ \cite{maiani_1,maiani_2} and a
$D^{*0}\!\!-\!\!\bar{D}^{0}$ molecule
\cite{molecule_Close,molecule_V,molecule_T} bound by pion exchange
were also considered.

In fact, $D^{}\!\!-\!\!\bar{D}^{}$ molecules were previously
predicted
\cite{pre_voloshin,pre_rujula,pre_t1,pre_ericson,pre_t2}.
T\"{o}rnqvist \cite{pre_t1,pre_t2} showed that one-pion exchange
potential is likely to bind a few states composed of two mesons
and referred to such deuteronlike meson-meson bound states as
deusons.

One of the motivations of the molecular interpretation of
$X(3872)$ is the observation that $M(X)$ equals, within errors,
$M(D^{*0})+M(\bar{D}^{0})$. Thus, the loosely bound molecule
$D^{*0}\!\!-\!\!\bar{D}^{0}$ is expected to have a large
$r_{r.m.s}$. Close and and Page \cite{molecule_Close} estimated
that $r_{r.m.s.} \approx 7 \ \rm{fm}$. They \cite{molecule_Close}
also pointed out that $D^{0}\!\!-\!\!\bar{D}^{0}$ molecule does
not exit because the $\pi D^0 \bar{D}^0$ vertex vanishes by parity
conservation.

The electromagnetic radiative decay channel $X\rightarrow D^0
\bar{D}^0 \gamma $ can be very useful in providing an insight into
the structure of the $X(3872)$. Voloshin \cite{molecule_V}
considered interference and binding effects in the radiative
decays of $X(3872)$ and pointed out that the molecular component
of $X(3872)$ ,
$\frac{1}{\sqrt{2}}\left[|D^{*0}\bar{D}^{0}\rangle\pm|D^0\bar{D}^{*0}\rangle\right]$
, can be revealed by a distinct pattern of interference between
the underlying decays of $D^{*0}$ and $\bar{D}^{*0}$.

It was Dicke \cite{Dicke} who first pointed out that the decay
rate of an excited atom is affected if a second ground state atom
is in its neighborhood. The decay rate can be enhanced up to
double that of an isolated atom (superradiance) if the two atoms
are in a symmetric state. If the two atoms are in an antisymmetric
state, the decay rate is suppressed (subradiance). The
superradiance and the subradiance factors are functions in the
distance separating the two atoms and approaches one  when the
distance between the atoms is large compared to the wavelength of
the emitted radiation. If the molecular interpretation of
$X(3872)$ is confirmed, this resonance can be a good laboratory
for investigating superradiance and subradiance.

In the present work, we evaluate the superradiance and the
subradiance factors for the electromagnetic radiative decay of
$X(3872)$ assuming that the molecular structure takes the form
$\frac{1}{\sqrt{2}}\left[|D^{*0}\bar{D}^{0}\rangle\pm|D^0\bar{D}^{*0}\rangle)\right]$
using the coordinate space representation. Our results are similar
to the results of Voloshin \cite{molecule_V} evaluated in the
momentum space representation. The coordinate space representation
is more suitable for clarifying the interference in the radiative
decay process.

In Sec. 2, we evaluate the electromagnetic radiative decay of the
$D^*$ meson using the Golden Rule. In Sec. 3, we consider the
electromagnetic radiative decay of $X(3872)$ and evaluate the
superradiance and the subradiance factors. In Sec 4, we present
the conclusions.

\section{Electromagnetic radiative decay of the $D^*$ meson \label{d}}

The width of the electromagnetic radiative decay channel of an
initial state $|i\rangle$ decaying into a final state $|f\rangle$
can be calculated using the Golden Rule
\begin{equation}
\Gamma=\sum_{\lambda}\int2\pi|\langle
f|H_{int}|i\rangle|^2\frac{V\omega^2d\Omega}{\hbar
c^3(2\pi)^3},\label{e_Golden_Rule}
\end{equation}
where $H_{int}$ is the electromagnetic interaction connecting the
initial and final states. We sum over the photon polarization
$\lambda$ and integrate over all photon directions.

The Hamiltonian of the $D$ meson , $D^*$ meson, and the
electromagnetic fields can be written as
\begin{equation}
H=H_M+H_F+H_{int}, \label{e_h_meson_field}
\end{equation}
where $H_M$ is the meson Hamiltonian whose lowest eigenstates are
$|D\rangle$ and $|D^*\rangle$, $H_F$ is the electromagnetic field
Hamiltonian, and $H_{int}$ is the interaction Hamiltonian between
the mesons and the electromagnetic fields. For the electromagnetic
radiative decay of the $|D^*\rangle$ state, $H_{int}$ takes the
form
\begin{equation}
H_{int}=-\vec{\mu}\cdot \vec{B}, \label{e_hmudotb}
\end{equation}
where $\vec{\mu}$ is the magnetic moment and the $\vec{B}$ is the
magnetic field. The magnetic field can be expanded in terms of the
annihilation and the creation operators $a(\vec{k},\lambda)$ and
$a^{\dag}(\vec{k},\lambda)$ in the usual form
\begin{equation}
\vec{B}(\vec{x})=i
\sum_{\vec{k},\lambda}\sqrt{\frac{2\pi\hbar\omega}{V}}
\hat{k}\times\vec{\varepsilon}(\vec{k},\lambda)
\left[a(\vec{k},\lambda)e^{i\vec{k}\cdot\vec{x}}-
a^{\dag}(\vec{k},\lambda)e^{-i\vec{k}\cdot\vec{x}}\right].
\label{e_B}
\end{equation}
The matrix element of the transition is given by
\begin{eqnarray}
\langle f|H_{int}|i\rangle & =& \langle D\gamma|H_{int}(\vec{x})
|D^*\rangle \nonumber\\
&=&i\sqrt\frac{2\pi\hbar\omega}{V}\langle D| \vec{\mu}|D^*\rangle
\cdot\hat{k}\times\vec{\varepsilon}(\vec{k},\lambda) e^{-i\vec{k}
\cdot \vec{x}}. \label{e_matrix_element_D}
\end{eqnarray}
%The electromagnetic partial width can be be written as
%\begin{equation}
%\gamma|H_{MF}|D^*\rangle|^2\frac{V\omega^2d\Omega}{\hbar
%c^3(2\pi)^3}\label{e_gammaD}
%\end{equation}
From Eq. (\ref{e_Golden_Rule})  and Eq. (\ref{e_matrix_element_D})
and using the relation
\begin{equation}
\sum_\lambda \varepsilon_i(\vec{k},\lambda)
\varepsilon_j(\vec{k},\lambda)=\delta_{ij}-\hat{k}_i\hat{k}_j,
\label{e_spin}
\end{equation}
we get
\begin{equation}
\Gamma_{D^*}=\int2\pi \frac{2\pi\hbar \omega}{V}\langle D|
\mu^i|D^*\rangle\langle D|\mu^j|D^*\rangle^*
 \left[\delta_{ij}-\hat{k}_i\hat{k}_j\right] \frac{V\omega^2d\Omega}{\hbar
 c^3(2\pi)^3}.
\label{e_gammaD2}
\end{equation}
Integrating over all angles of the emitted photon, we get
\begin{equation}
\Gamma_{D^*}=\frac{4}{3}\frac{\omega^3}{c^3}|\langle
D|\vec{\mu}|D^*\rangle |^2.
 \label{e_gammaDf}
\end{equation}

\section{Electromagnetic radiative decay of $X(3872)$\label{x}}

Assuming the validity of the molecular interpretation of
$X(3872)$, we can write the Hamiltonian of $X(3872)$ and the
electromagnetic fields in the form
\begin{equation}
H=H_1+H_2+H_{12}+H_F+H_{int}, \label{e_h_molecule}
\end{equation}
where $H_{1(2)}$ is the Hamiltonian of the first (second) meson,
$H_{12}$ is the Hamiltonian describing the relative motion of the
two mesons in the molecule, $H_F$ is Hamiltonian of the
electromagnetic fields, and $H_{int}$ is the Hamiltonian of the
interaction between the first and the second mesons with the
electromagnetic fields. The lowest eigenstates of $H_1$ are
$|D\rangle$ and $|D^*\rangle$, while the lowest eigenstates of
$H_2$ are $|\bar{D}\rangle$ and $|\bar{D}^*\rangle$.

The initial state can be written as
$|i\rangle=\frac{1}{\sqrt{2}}\left[|D^*\bar{D}\rangle\pm|D\bar{D}^*\rangle\right]$
while the final state is $|f\rangle=|D\bar{D}\gamma\rangle$, and
the interation Hamiltonian,
$H_{int}=H_{int}(\vec{r_1})+H_{int}(\vec{r_2})$, is evaluated at
the positions of the first and the second mesons $\vec{r_1},
\vec{r_2}$
\begin{eqnarray}
\vec{r_1}&=&\vec{R}-\frac{M_2}{M_1+M_2}\vec{r},\nonumber\\
\vec{r_2}&=&\vec{R}+\frac{M_1}{M_1+M_2}\vec{r},
 \label{e_r}
\end{eqnarray}
where $\vec{R}$ is the center of mass of the molecule and
$\vec{r}$ is the relative coordinate from the first to the second
meson. In the center of mass system, $\vec{R}=0$, and the matrix
element of the transition $\langle f|H_{int}|i\rangle$ is given by
\begin{eqnarray}
\langle f|H_{int}|i\rangle \!\!\!\!\!& =& \!\!\!\!\!\langle
D\bar{D}\gamma| \left \{ H_{int}(\vec{r}_1)+H_{int}(\vec{r}_2)
\right \}
\frac{1}{\sqrt{2}}\left[|D^*\bar{D}\rangle\pm|D\bar{D}^*\rangle\right] \nonumber\\
\!\!\!\!\!&=&\!\!\!\!\!\frac{i}{\sqrt{2}}\sqrt\frac{2\pi\hbar\omega}{V}
\left\{\langle D| \vec{\mu}|D^*\rangle
e^{+i\frac{M_2}{M_1+M_2}\vec{k} \cdot \vec{r}} \pm \langle
\bar{D}| \vec{\mu}|\bar{D}^*\rangle
e^{-i\frac{M_1}{M_1+M_2}\vec{k} \cdot
\vec{r}}\right\}\cdot\hat{k}\times\vec{\varepsilon}(\vec{k},\lambda).
\label{e_h_int1}
\end{eqnarray}
Since $\langle \bar{D}| \mu^i|\bar{D}^ *\rangle = \eta \langle D|
\mu^i|D^*\rangle$, where $\eta=-1$, we get
\begin{eqnarray}
|\langle f|H_{int}|i\rangle|^2
\!\!\!&=&\!\!\!\frac{1}{2}\frac{2\pi\hbar\omega}{V} \langle D|
\mu^i|D^*\rangle \langle D|
\mu^j|D^*\rangle^*\varepsilon_i(\vec{k},\lambda)
\varepsilon_j(\vec{k},\lambda)2\left\{1 \pm \eta
\cos\vec{k}\cdot\vec{r}\right\}. \label{e_h_int2}
\end{eqnarray}
The electromagnetic radiative decay width of the molecular state
$X$ of the two mesons separated by a distance $r$ is given by
\begin{equation}
\Gamma^{\pm}_{X}(r)=\int2\pi \frac{2\pi\hbar \omega}{V}\langle D|
\mu^i|D^*\rangle\langle D|\mu^j|D^*\rangle^*
 \left[\delta_{ij}-\hat{k}_i\hat{k}_j\right]\left\{1 \pm
\eta \cos\vec{k}\cdot\vec{r}\right\} \frac{V\omega^2d\Omega}{\hbar
c^3(2\pi)^3}.
 \label{e_gammaxr1}
\end{equation}
To evaluate the angular integral, we write
\begin{equation}
\int \left[\delta_{ij}-\hat{k}_i\hat{k}_j\right]\left\{1 \pm \eta
\cos
\vec{k}\cdot\vec{r}\right\}d\Omega=2\pi\left[\delta_{ij}A_{\pm} +
\hat{r}_i\hat{r}_jB_{\pm}\right].
 \label{e_AB1}
\end{equation}
Multiplying both sides by $\delta_{ij}$ and $\hat{r}_i\hat{r}_j$
and summing over $i$ and $j$ we get two equation, which are solved
for $A_{\pm}$ and $B_{\pm}$
\begin{eqnarray}
A_{\pm}&=&\frac{4}{3}\pm 2 \eta \left\{ \frac{\sin k r}{k r} +
\frac{\cos k r}{(k r)^2}
- \frac{\sin k r}{(k r)^3} \right\},\nonumber\\
B_{\pm}&=&\pm 2 \eta \left\{ -\frac{\sin k r}{k r} -3 \frac{\cos k
r}{(k r)^2} +3 \frac{\sin k r}{(k r)^3}\right\}. \label{e_AB2}
\end{eqnarray}
From Eqs. (\ref{e_gammaxr1},\ref{e_AB1},\ref{e_AB2}), we get
\begin{equation}
\Gamma^{\pm}_{X}(r)=\frac{\omega^3}{c^3}|\langle
D|\vec{\mu}|D^*\rangle |^2\left[A_{\pm}+\cos^2\theta
B_{\pm}\right], \label{e_gammaxr2}
\end{equation}
where $\theta$ is the angle between $\vec{\mu}$ and $\vec{r}$. For
arbitrary directions of the magnetic moment, we take the average
value of $\cos^2\theta = \frac{1}{3}$, and we get
\begin{equation}
\Gamma^{\pm}_{X}(r)=\frac{\omega^3}{c^3}|\langle
D|\vec{\mu}|D^*\rangle |^2\frac{4}{3}\left[1\pm \eta \frac{\sin k
r}{k r}\right]. \label{e_gammaxr3}
\end{equation}
Using  Eq. (\ref{e_gammaDf}) for the width of the $D^*$ meson, and
noting that in our case $\eta=-1$, we can write Eq.
(\ref{e_gammaxr3}) in the form
\begin{equation}
\Gamma^{\pm}_{X}(r)=\Gamma_{D^*}S^{\mp}(r), \label{e_gammaS1}
\end{equation}
where $S^{\pm}(r)$ is given by
\begin{equation}
S^{\pm}(r)=\left[1\pm  \frac{\sin k r}{k r}\right].
\label{e_gammaS2}
\end{equation}
We can call $S^+(r)$ and $S^-(r)$ the superradiance and the
subradiance factors when the two radiating mesons are separated by
a distance $r$. These factors describe the effect of the
interference between the two radiating sources $D^*$ and
$\bar{D}^*$ on the decay width as a function of $k r$. We notice
that since $\langle \bar{D}| \mu^i|\bar{D}^ *\rangle = \eta
\langle D| \mu^i|D^*\rangle$, where $\eta=-1$, the symmetric state
is subraiant while the antisymmetric state is suprradiant, unlike
the situation usually encountered in atomic physics where the
symmetric state is superradiant and the antisymmetric state is
subradiant.

\begin{figure}[h!tb]
\centerline{\epsfxsize=1.0\textwidth\epsfbox{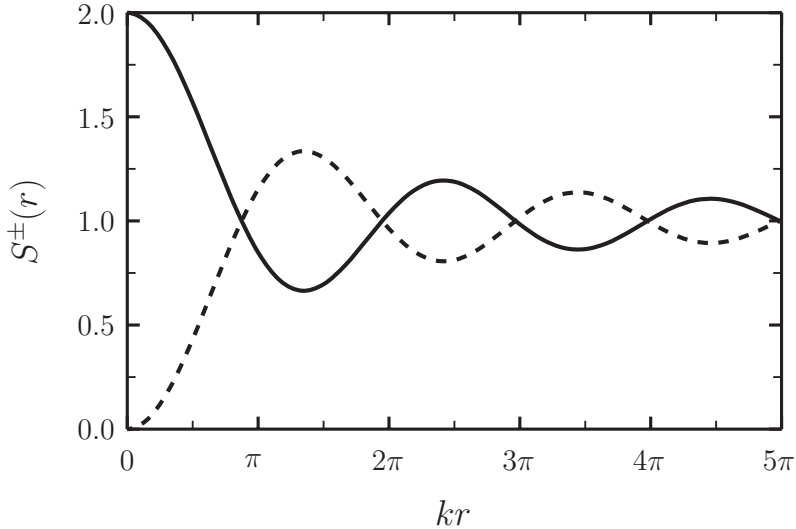}} \caption{
\vspace{0.0cm} The superradiance factor $S^+(r)$ (solid line) and
the subradiance factor $S^-(r)$(dashed line) against $k r$.}
\label{fig_spmkr}
\end{figure}

Fig. \ref{fig_spmkr} shows the dependence of the superradiace
factor $S^+(r)$ and the subradiance factor $S^-(r)$ on $k r$. For
$r=0$ the the decay width is doubled in the superradiant state
while the subradiant state is stable. As the distance between the
two radiating mesons increases, both the superradiace and the
subradiance factors approach one.
%fig_spmkr

The distance between the two mesons is not fixed but varies
according to the molecular wavefunction which results from the
solution of Schr\"{o}dinger Equation for the Hamiltonian $H_{12}$
of Eq. (\ref{e_h_molecule}). We follow Voloshin \cite{molecule_V}
and consider the molecular wavefunction as an $S$-state in the
form

\begin{equation}
\psi(\vec{r})=\sqrt{\frac{1}{2\pi r_0}}\frac{e^{-r/r_0}}{r},
\label{e_psi}
\end{equation}
where $r_0=1/\sqrt{2\mu w}$,
$\mu=[M(D^*)M(\bar{D})]/[M(D^*)+M(\bar{D})]$ is the reduced mass,
and $w=M(D^*)+M(\bar{D})-M(X)$ is the binding energy of the
molecule. Wavefunctions of nonzero values of orbital angular
momentum can also be treated similarly.

The electromagnetic radiative decay width of the molecular state
of $X(3872)$ can be evaluated by taking the average of
$\Gamma^{\pm}_X(r)$ in the molecular state $\psi(\vec{r})$
considered in Eq. (\ref{e_psi}), thus we get
\begin{eqnarray}
\Gamma^{\pm}_{X}=\int
\psi(\vec{r})^*\Gamma^{\pm}_X(r)\psi(\vec{r})4\pi r^2dr,
\label{e_gammax2}
\end{eqnarray}
which is easily evaluated, and we can write
\begin{eqnarray}
\Gamma^{\pm}_{X}=\Gamma_{D^*}S^{\mp}, \label{e_gammaw1}
\end{eqnarray}
where
\begin{eqnarray}
S^{\pm}=\left[1\pm \frac{2}{k r_0}\arctan\frac{k r_0}{2}\right].
\label{e_gammaw2}
\end{eqnarray}
This is our final result, and it is in agreement  with the result
obtained by Voloshin \cite{molecule_V} using the momentum space
representation in the calculation of the interference and the
molecular wavefunction.

\begin{figure}[h!tb]
\centerline{\epsfxsize=1.0\textwidth\epsfbox{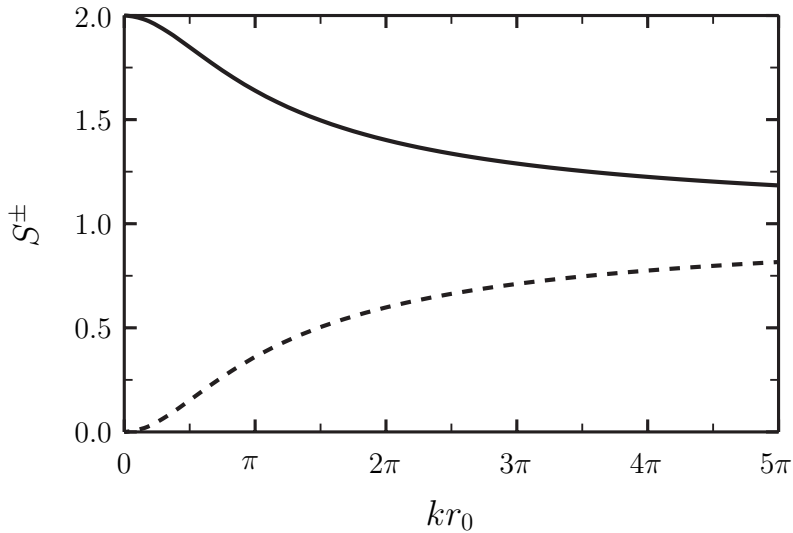}} \caption{
\vspace{0.0cm} The superradiance factor $S^+$ (solid line) and the
subradiance factor $S^-$(dashed line) against $k r_0$.}
\label{fig_spm}
\end{figure}

Fig. \ref{fig_spm} shows the dependence of the superradiace factor
$S^+$ and the subradiance factor $S^-$ on $k r_0$. For small $k
r_0$ the superradiant limit of 2 and the subradiant limit of zero
are reached, while for large $k r_0$ both factors approach one. It
is interesting to note that while $S^{\pm}(r)$ oscillate about the
value of 1 as shown in Fig. \ref{fig_spmkr}, $S^{+}$ is
monotonically decreasing to 1 and $S^{-}$ is monotonically
increasing to 1.
%
%fig_spm

For the decay of $X(3872)$ into $D^0 \bar{D}^0\gamma$,
$k=[M(D^{*0})^2-M(D^0)^2]/2M(D^{*0})=137 \ \rm{MeV}$ and
$r_0=1/\sqrt{2\mu w}=0.023 \ \rm{MeV}^{-1}$ for $\mu=966 \
\rm{MeV}$ and $w=1 \ \rm{MeV}$, so that $k r_0=3.1$. For these
numerical values, $S^+=1.64$ and $S^-=0.36$.

Measurements of the radiative decay width and comparison with Fig.
\ref{fig_spm} can provide valuable information about the structure
and the dynamics of the state $X(3872)$. The molecular size, the
binding energy, and the symmetry of $X(3872)$ can be studied in
this way. This state can be a good laboratory for investigating
the superradiance and the subradiance phenomena. In addition to
the neutral channel $D^0\bar{D}^0\gamma$, the charged channel
$D^+\bar{D}^-\gamma$, although having a smaller width, will have
larger binding energy $w$, smaller $r_0$ and consequently stronger
interference effects.
%\begin{center}
\section{Conclusions\label{conclusions}}
%\end{center}

The narrow resonance $X(3872)$ confirmed by many experiments may
have a large component of its wave function as $D^* \bar{D}$ and
$D \bar{D}^*$ molecule. The electromagnetic decays of the $D^*$
and $\bar{D}^*$ mesons will reveal interference effects which
depend on the distance between these two mesons and the wavelength
of the emitted photon as pointed out by Dicke. Insight into the
internal structure and the dynamics of $X(3872)$ can be gained by
studying the electromagnetic radiative decay width into neural and
charged $D$ and $\bar{D}$ mesons. Hadronic molecules are
interesting in themselves and the phenomena of superradiance and
subradiance may help reveal their features.

{\bf Acknowledgments}

I thank Prof. J.P. Vary for a critical reading of the manuscript.

%PLB
\newpage

%PRD

\end{document}